\documentclass[preprint,3p,twocolumn]{elsarticle}
\usepackage{graphicx,caption}

\usepackage{url}            
\usepackage{subfiles}
\usepackage{array}
\newcolumntype{P}[1]{>{\centering\arraybackslash}m{#1}}
\usepackage{lineno}
\usepackage{microtype}

\journal{Computer Physics Communications}

\begin{document}

\begin{frontmatter}

\title{Dynamic load balancing with enhanced shared-memory parallelism for particle-in-cell codes}

\author[1]{Kyle G. Miller}
\ead{kylemiller@physics.ucla.edu}

\author[1]{Roman P. Lee\corref{cor1}}
\ead{romanlee@physics.ucla.edu}
\cortext[cor1]{Corresponding author}

\author[1]{Adam Tableman}

\author[2]{Anton Helm}

\author[3,2]{Ricardo A. Fonseca}
\ead{ricardo.fonseca@iscte-iul.pt}

\author[1]{Viktor K. Decyk}

\author[1]{Warren B. Mori}
\ead{mori@physics.ucla.edu}

\address[1]{Department of Physics and Astronomy, University of California, Los Angeles, California 90095, USA}
\address[2]{GoLP/Instituto de Plasmas e Fus\~{a}o Nuclear, Instituto Superior T\'{e}cnico, 1049-001 Lisboa, Portugal}
\address[3]{ISCTE - Instituto Universit\'{a}rio de Lisboa, Av. For\c{c}as Armadas, 1649-026 Lisboa, Portugal}

\begin{abstract}
Furthering our understanding of many of today's interesting problems in plasma physics---including plasma based acceleration and magnetic reconnection with pair production due to quantum electrodynamic effects---requires large-scale kinetic simulations using particle-in-cell (PIC) codes. However, these simulations are extremely demanding, requiring that contemporary PIC codes be designed to efficiently use a new fleet of exascale computing architectures. To this end, the key issue of parallel load balance across computational nodes must be addressed.  We discuss the implementation of dynamic load balancing by dividing the simulation space into many small, self-contained regions or ``tiles,'' along with shared-memory (e.g., OpenMP) parallelism both over many tiles and within single tiles.  The load balancing algorithm can be used with three different topologies, including two space-filling curves.  We tested this implementation in the code \textsc{Osiris} and show low overhead and improved scalability with OpenMP thread number on simulations with both uniform load and severe load imbalance.  Compared to other load-balancing techniques, our algorithm gives order-of-magnitude improvement in parallel scalability for simulations with severe load imbalance issues.
\end{abstract}


\end{frontmatter}


\section{Introduction}

The particle-in-cell (PIC) algorithm is widely used to study interesting problems where discrete particles or agents interact through fields. The PIC algorithm has thus been widely used in the kinetic modeling of plasmas, where the fields can either be electrostatic or electromagnetic.  However, the nature of tracking individual particles over long periods of time makes the PIC algorithm computationally expensive, requiring the use of large-scale high-performance computing (HPC) resources. With the advent of exascale computing, HPC architectures are undergoing rapid change; since 2004, clock rates have stabilized and growth on top-ranked systems has come almost entirely from increased parallelism. The PIC algorithm is sensitive to load imbalance on excessively parallel machines because simulation particles move about and may accumulate on a fraction of the computing resources. As the scale of massively parallel computing architectures continues to intensify, the final push toward exascale and beyond will require significant adaptation of software to take advantage of the increased parallelism available in the hardware.

Distributing computational load evenly across resources can be achieved through multiple levels of parallelism. At the highest level one can parallelize across distributed-memory processing elements (PEs). Parallelism on this level is often implemented via the Message Passing Interface (MPI), and for clarity, in this paper PE always refers to an MPI process. In addition, the increasing number of shared-memory CPUs inside compute nodes, as well as the increasing number of cores inside today's CPUs, allows for parallelism within a processing element (e.g., via OpenMP or Pthreads). Finally, the use of many-core accelerators such as Graphical Processing Units (GPUs) also allows for parallelism via CUDA or OpenACC.



In this paper we present developments to improve the parallel scalability of the particle-in-cell (PIC) code \textsc{Osiris}~\cite{Fonseca2002OSIRIS:Accelerators,Fonseca2013}, which can be applied to any massively parallel PIC code. On top of the parallel computing challenges presented by evolving hardware architectures, parallelizing a PIC code while maintaining load balance is inherently challenging on an algorithmic level. A PIC code contains two main data structures that comprise the computational load in a simulation: particles that can occupy any position in the simulation domain and field quantities that are discretized on a mesh grid. Parallelization is done by distributing particles and grid points among PEs, with load balance being achieved when the computational load, here defined as the calculation time, associated with these structures is distributed uniformly. \textsc{Osiris}~\cite{Fonseca2002OSIRIS:Accelerators,Fonseca2013} and most cutting-edge PIC codes use a spatial grid-based domain decomposition \cite{Derouillat2018,Bowers2008UltrahighSimulation,Grote2005TheBeams}, where each PE is responsible for a subset of the global spatial grid as well as any particles located there. While this domain decomposition algorithm is widely used, it suffers from the possibility that a large number of particles may move into a single PE (as is often the case for simulations of plasma-based acceleration, laser-solid interactions, or magnetic reconnection with pair production). Thus maintaining acceptable load balance can be challenging, given that the computational load will generally scale with the number of particles.

Various strategies have been implemented in an effort to dynamically load balance PIC codes using grid-based decomposition.
Perhaps the most straightforward load balancing scheme is to enlarge distributed-memory spatial domains by using a large number of shared-memory cores on each PE and distribute the computational load evenly across these cores. This allows for localized load-imbalance situations, such as large density spikes, to be smoothed out, generally leading to good improvements in performance and scalability~\cite{Fonseca2013}. This offers limited relief, however, due to hardware limitations on the available number of cores.
Another solution based on a static equidistant domain decomposition is known as the taskfarm alternative~\cite{Othmer2002}. Here, each regular and equally sized distributed-memory sub-domain is further subdivided uniformly into tasks, with particles sorted accordingly. To process particles, each PE works serially through its own set of tasks before accessing, completing and returning the tasks of other PEs with higher load. If tasks are small enough, load is balanced since no PE is ever idly waiting for remaining tasks to be completed.
A different approach is taken by Liewer and Decyk who pioneered the idea of shifting PE boundaries in their 1988 algorithm GCPIC~\cite{Ferraro1993,Liewer1989}. Computational load is projected onto one axis so that the problem of load balancing becomes one dimensional. Partitions are found along this axis, and the resulting domains are partitioned in the second and third dimensions in the same way. Similar approaches based on rectilinear partitioning are taken in~\cite{Saule2012,Fonseca2013,Surmin2015}.
Dynamic load balancing can also be achieved by decomposing the simulation into many small units called tiles (or patches). Each PE handles one or more of these units, with the algorithm dynamically assigning them between PEs to maintain an even load~\cite{Germaschewski2016,Derouillat2018}. Similar strategies have also been employed on GPU architectures~\cite{Decyk2014}.


In this paper we extend the previous dynamic load balancing algorithm of \textsc{Osiris}~\cite{Fonseca2013} by dividing the global simulation space into many small, self-contained ``tiles,'' which contain all particle and grid quantities for a particular region of space. One or more threads are assigned to each tile depending on its computational load, after which the parallel domain decomposition is determined by assigning one or more tiles to each PE such that computational load is as balanced as possible. The ability to assign multiple threads to each tile---following the shared-memory parallelization algorithm already present in \textsc{Osiris}---allows us to significantly improve performance for simulations with small regions of high particle density by enabling the parallel use of a multi-core PE on a single tile. This provides significant improvement over previous tile-based dynamic load balancing implementations~\cite{Germaschewski2016,Derouillat2018} that allow for only one thread per tile on CPUs. We find that this feature allows for the use of larger tiles---reducing the overhead of passing particles between tiles---while still maintaining load balance. Our implementation also scales well with thread number and gives particularly large speedups for very imbalanced simulations, being well suited for efficient use in today's evolving HPC climate as available on-chip thread count continues to climb.

This paper is organized as follows: In Sec.~\ref{sec:methodology}, we discuss the implementation of the tile structure into \textsc{Osiris}, including both distributed-memory and shared-memory parallelization schemes. We discuss the overhead and performance of the tiling scheme in Sec.~\ref{sec:results} by analyzing simulations both with and without load imbalance, including a 3-D simulation of particle wakefield acceleration. Compared to previous \textsc{Osiris} algorithms, our tile-based implementation of dynamic load balancing gives an order-of-magnitude increase in scalability with thread number and more than a factor of 2 overall speedup for two different physics simulations.

\section{Methodology} \label{sec:methodology}

In the last two decades, the continued growth of top-ranked HPC systems has come almost entirely from increased parallelism, with present systems comprising up to $\sim 10^6$ cores.  At the highest level of parallelism, these massive computer systems are viewed as a network of distributed-memory PEs amongst which the simulation can be partitioned. Given that these PEs do not share memory, using a spatial domain decomposition requires that particle and field data be exchanged between neighboring PEs at each time step. Domains for each PE should be structured such that the computation time on a given region is much larger than the time spent communicating boundary information, i.e. maximizing the computational volume to boundary surface area ratio to minimize parallelization overhead.

Previously, the domain decomposition in \textsc{Osiris} was structured such that each PE had only one neighboring PE in each direction (i.e., domain corners always matched up).  These domains could be statically assigned at the beginning of the simulation or changed dynamically throughout to maintain load balance~\cite{Fonseca2013}. Aligning domain corners simplified the communication pattern for the sharing of boundary information, but limited the achievable load balance.  In an effort to improve dynamic load balancing and to enhance shared-memory parallelism in \textsc{Osiris}, we decompose the global simulation space into many small, static, regularly spaced rectangular regions called ``tiles'' which have aligned corners, following a strategy similar to \cite{Germaschewski2016,Derouillat2018}. The domain decomposition is then determined by assigning a collection of one or more tiles to each PE such that computational load is balanced. These tiles can be exchanged dynamically between all PEs at chosen intervals throughout the simulation to maintain load balance.

\begin{subsection}{Distributed-memory parallelism}

To determine the most balanced parallel partition, i.e., assignment of tiles to PEs, we must first devise a way to quantify and organize the computational work required to process each tile. To this end, we create a load array with the same dimensions as the simulation and containing one entry per tile. Assuming that the computational load scales linearly with both the number of particles and the number of grid points\footnote{This is accurate for finite-difference-based field solvers. For other types of field solvers a different load formula can be straightforwardly derived.}, we compute the array entry for the $i^{\mathrm{th}}$ tile, $\mathcal{L}_i$, as $\mathcal{L}_i = \mathcal{N}_{i,\mathrm{part}} +\mathcal{C} N_{i,\mathrm{cells}}$ as in \cite{Germaschewski2016}, where $\mathcal{N}_{i,\mathrm{part}}$ and $\mathcal{N}_{i,\mathrm{cells}}$ are the number of particles and cells in the $i^{\mathrm{th}}$ tile, respectively, and the cell weight $C$ represents the computational load of a single cell compared to one particle. This depends on simulation parameters such as interpolation level or field solver type, and is left for the user to define. We found that for typical 2- and 3-D simulations $C\approx~0.5\textrm{--}2$ is effective. The load array can then be partitioned into $\mathcal{N}_{\mathrm{PE}}$ regions of contiguous tiles such that the total load in each partition is as close as possible to the ideal average load, $\sum_{i} \mathcal{L}_i/\mathcal{N}_{\mathrm{PE}}$, where $\mathcal{N}_{\mathrm{PE}}$ is the total number of PEs. Partitioning a 1-D load array in this way is trivial since boundaries must be determined only along one dimension.  However, partitioning a 2- or 3-D load array is more complex, and multiple solutions may be considered.

One approach proposed by Saule et al.~\cite{Saule2012} is to consider 2-D partitions where each PE is left with a rectangular region of space, with each boundary connecting to one or more neighbors. Partitions where each boundary connects to a single neighbor (matching corners)---referred to as a rectilinear partition in~\cite{Saule2012}---was previously implemented in \textsc{Osiris}~\cite{Fonseca2013}. A more complex scheme, referred to as $P\times Q$ jagged in~\cite{Saule2012}, allows for multiple neighbours along boundaries in the $x$ dimension, and only one neighbor in $y$. This allows for decomposing the multi-dimensional load balance problem into separate uni-dimensional scenarios: we first define $P$ partitions along $x$ so that the load is evenly divided among them; we then proceed by dividing each of these $P$ partitions into $Q$ sub-partitions such that the computational load for all $Q$ sub-partitions in a $P$ partition is equal. This process can be straightforwardly extended to 3 dimensions. The $P\times Q$ jagged partition is intuitive and has rather simple boundaries between PEs, but for some computational load distributions it may not yield a perfectly balanced configuration, given that the boundary positions are limited to grid cell boundaries. See Ref.~\cite{Liewer1989} for an example of a PIC code using the jagged partition.

Another method to load balance in multiple dimensions is to place the tiles sequentially along a space-filling curve \cite{Germaschewski2016}. After estimating the optimal load per PE, we start with the first PE and, following the curve, assign tiles to it until the load is close to the optimal value. We then proceed to the next PE/tile until all tiles have been assigned. The advantage of using a space-filling curve for ordering the tiles as opposed to, say, choosing the tiles based solely on their computational load, is that we will end up with simulation domains that are contiguous in space and have relatively simple boundaries with a small number of neighbors. We implement this method by creating a 1-D load array, where tiles are ordered in the array using their position along the space-filling curve, then dividing this array into $\mathcal{N}_{\mathrm{PE}}$ segments with roughly equal load. Tiles falling on each segment will be assigned to the corresponding PE.

\begin{figure}[t]
    \centering

    \captionsetup{width=1.0\linewidth}
    \includegraphics[width=0.95\linewidth]{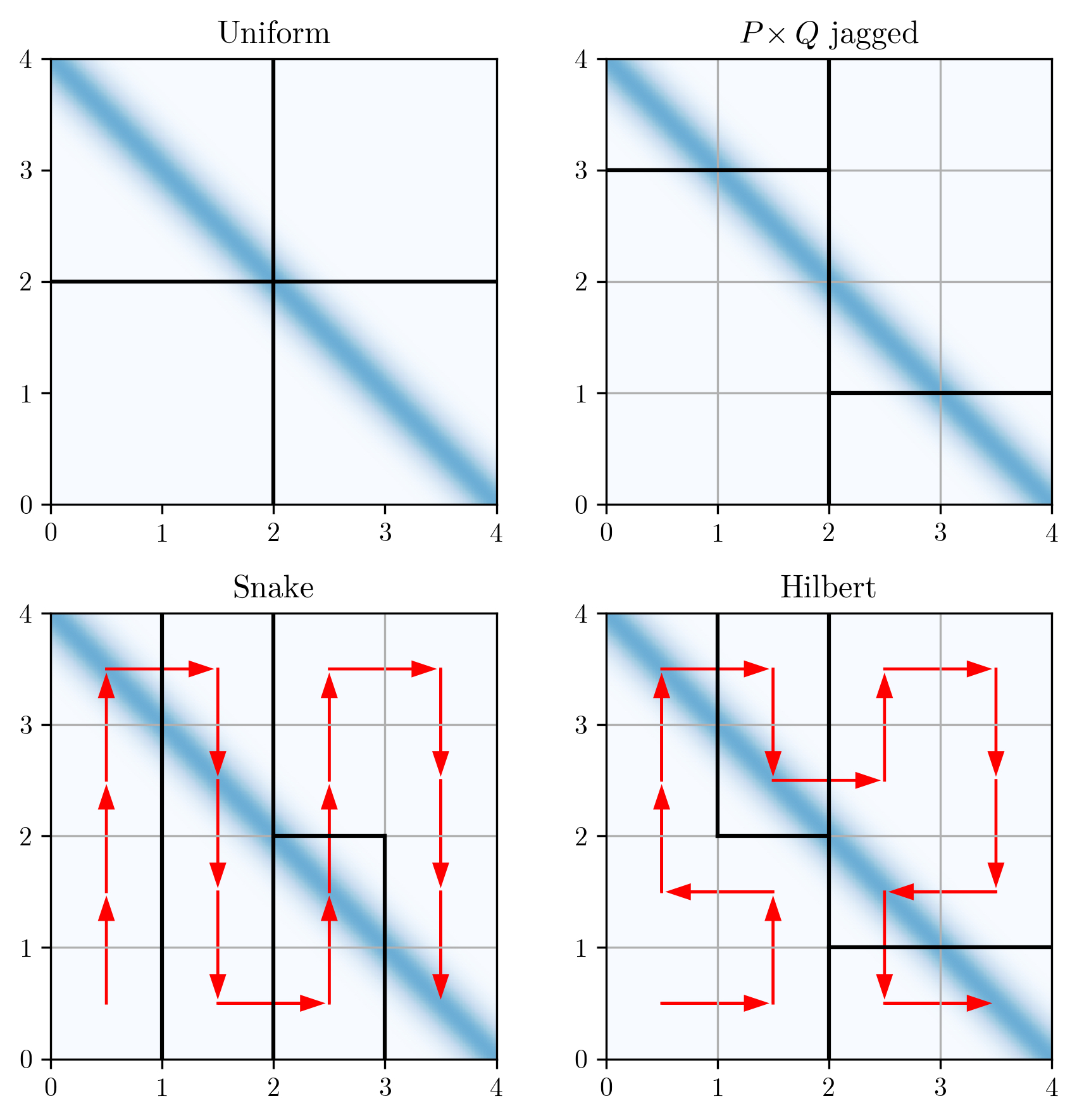}
    \caption{Domain decomposition using 4~PEs for a high-density diagonal stripe of particles (blue).  We show a uniform partition without tiles, as well as partitions using the three load balancing schemes implemented here.  Gray lines indicate tile boundaries, black lines indicate PE (i.e., MPI) boundaries, and red arrows trace out tile ordering along the space-filling curve.  Note that with uniform partitioning, two PEs contain very few particles.  For the density profile and tile size shown here, all three tiles schemes achieve roughly the same degree of load balance.}
    \label{fig:dlb-schemes}
\end{figure}

We implemented both the $P\times Q$ jagged partition and the space-filling curve methods in \textsc{Osiris}, with two choices for the space-filling curve. The first choice (Snake) simply passes through each tile by snaking back and forth in simulation space.  This curve has no restrictions on domain size and yields the simplest boundaries, but does not maximize the ratio between computational volume and boundary surface area.  The second choice (Hilbert) traces through the tiles using a Hilbert curve \cite{Hilbert1891UeberSurface}, will maximize the ratio between computational volume and boundary surface area, but requires that the tile number be a power of 2 in the smallest dimension (and integer multiples of that number in other dimensions) and leads to more complex boundary shapes, potentially with more neighbors. Figure~\ref{fig:dlb-schemes} shows schematics of various domain decompositions using 4~PEs for a simulation featuring a high-density diagonal stripe of particles (blue) surrounded by vacuum.  We show a uniform partition without using tiles, as well as partitions using the three load balancing schemes implemented here.  Note that with uniform partitioning, two PEs contain very few particles.  For the density profile and tile size shown here, all three tile schemes achieve roughly the same degree of load balance across PEs, though for more complicated profiles the $P \times Q$ jagged scheme will usually not load balance as well as the Snake/Hilbert schemes. To minimize communication overhead, we group messages from multiple tiles with the same destination PE into a single message, which greatly improves MPI performance.

To summarize, load balancing is performed by calculating the load array, determining the load-balanced domain decomposition, and then distributing the tiles among PEs accordingly. This can be done solely at the beginning of the simulation or dynamically throughout the simulation at chosen intervals using current simulation information. The domain decomposition defines a mapping between tiles and PEs, which is used to determine which tiles are to be sent and received, where tiles with the same destination PE are grouped into a single MPI buffer to minimize the effect of latency. Note that each successive domain decomposition is independent of the previous decomposition, i.e., a single PE may send all of its tiles to various other PEs and receive entirely new ones.

\end{subsection}

\begin{subsection}{Shared-memory parallelism} \label{sec:shared-memory}

The parallelization and load balancing algorithm described in the previous section works well for several problems, but does not guarantee ideal load balance for challenging scenarios where a large number of particles accumulate in a small number of tiles. Consider such a situation, where the load on a particular tile corresponds to $\alpha$ times the optimal load per PE, where $\alpha>1$. In this case, the best the load balancing algorithm can do is to assign that tile to a single PE. However, the load for that PE will be $\alpha$ times the optimal value, and we will experience a slowdown by a factor of $\alpha$. To address these situations and further improve parallel performance, we exploit the possibility for shared-memory parallelism that is available on most of today's computer systems, where a single PE may have many cores or threads that share memory with one another.  While load balancing across distributed memory must be done with the coarse resolution of one tile, load balancing within a single PE with many threads can be done to greater resolution, for example, by dividing particles evenly among shared-memory threads.

For the case where multiple tiles are assigned to a single PE, there are two common approaches for processing the tiles using shared-memory parallelism. If there are many more tiles than available threads on the PE, load balance can be achieved by looping over tiles using a dynamic scheduler ("first come, first served"), with one thread per tile. This has been previously demonstrated by \cite{Derouillat2018}. However, if one single tile contains a very large number of particles, the associated thread will take much longer than the others and load balance will fail.

Alternatively, tiles can be processed in serial with all threads working on a single tile at any given time.  This way a single thread is never stuck on a tile with many particles.  Dividing up particles in a single region of space amongst shared-memory resources is a strategy commonly implemented in GPU codes~\cite{Decyk2014,Kong2011Particle-in-cellUnits}. However, given that all threads are assigned to the same region of space (tile), we must avoid memory collisions when doing the current deposition. This can be achieved either by using atomic operations, or by creating a separate electric current grid for each thread, then summing these arrays together after the current deposition is complete. Both options have specific drawbacks that impede parallel scalability for large numbers of threads, but the latter option generally gives better performance on CPUs and is used in \textsc{Osiris}~\cite{Fonseca2013}.



We take a novel approach and include both types of shared-memory parallelism in a single framework. For a given PE with $\mathcal{T}$ threads and total load $\mathcal{L}$, let $\mathcal{L}_j$ be the load found on its $j^{\mathrm{th}}$ tile. Each time step we calculate which tiles should be processed in parallel (one thread per tile) and which tiles should be processed in serial ($\mathcal{T}$ threads per tile). A tile is deemed ``heavy'' if its load is greater than the average load per thread, i.e., if $\mathcal{L}_j \geq \mathcal{L}/\mathcal{T}$, and ``light'' otherwise. However, if a PE has fewer tiles than threads, all tiles are classified as ``heavy'' to avoid idle threads. For routines like those in particle processing, light tiles are processed first by assigning one thread per tile using a dynamic scheduler (we use OpenMP). Once all light tiles are completed, each heavy tile is executed one at a time, with all threads processing particles on that tile in parallel.

A schematic of the heavy-light tile organization is shown in Fig.~\ref{fig:heavy-light} for a PE with five light tiles, two heavy tiles, and four threads. Using light tiles avoids data dependency issues when there are many more tiles than threads. The inclusion of heavy tiles ensures that no threads are idle in cases with (1) fewer tiles than threads on a PE or (2) disproportionately large load on a single tile. This implementation is critical for simulations with very high particle densities, such as plasma wakefield or shock simulations, where the distributed-memory load balance may result in PEs having just one tile containing a majority of particles.

\begin{figure}[t]
    \centering
    \captionsetup{width=1.0\linewidth}
    \includegraphics[width=1.0\linewidth]{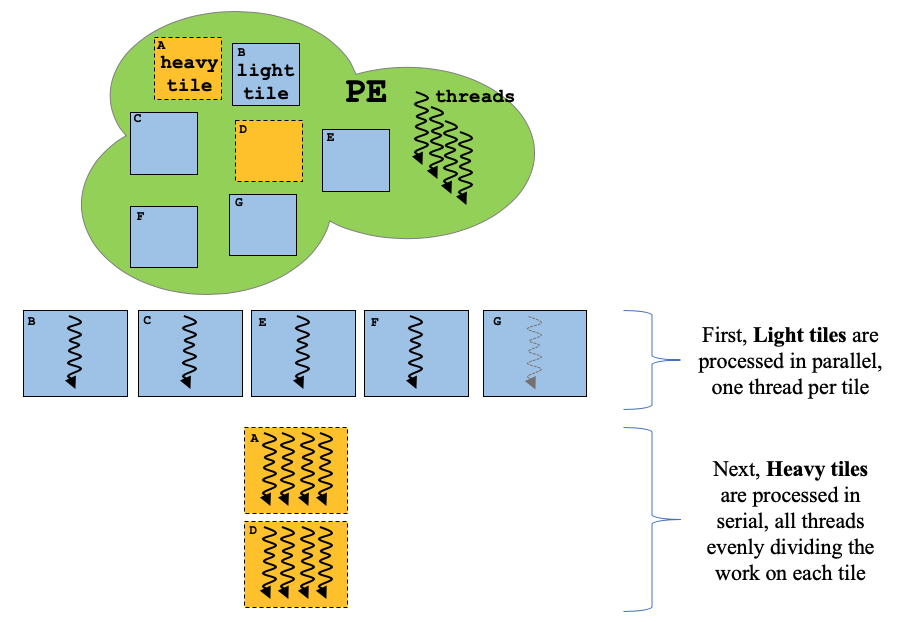}
    \caption{Work flow for shared-memory parallelization on a single MPI process. Each MPI process handles its tiles differently depending on their computational weight---a feature unique to our implementation. Using shared-memory threads (OpenMP in our case), tiles with load less than the average load per thread on that MPI process (light tiles) are processed in parallel with one thread per tile. Then tiles with above-average load (heavy tiles) are processed in serial with all threads dividing work on that tile evenly.}
    \label{fig:heavy-light}
\end{figure}

Alternatively, it is possible to use a task-based approach to thread over tiles. The algorithm begins by creating one task per tile, for all tiles. Light tiles will then be processed without further parallelism, while heavy tiles will be processed using multiple threads per tile, with the number of threads being proportional to the number of particles in the tile. The number of threads for each heavy tile can be calculated as $\mathcal{T}_i =  \lfloor \mathcal{T}\times(\mathcal{N}_i/\mathcal{N}) \rceil$ (round to nearest integer), where $\mathcal{T}$ is the total number of available threads in the PE, $\mathcal{N}$ is the total number of particles in the PE, and $\mathcal{N}_i$ is the number of particles in the tile. These tasks will be scheduled dynamically, assigning, if possible, a higher priority to light tiles. This approach should, in most scenarios, reduce the number of threads assigned to individual heavy tiles and ensure better load distribution across cores inside a PE, and will be further explored in a future publication.

\end{subsection}

\begin{subsection}{Tile boundary management} \label{sec:ub}

When using a parallelization scheme based on spatial domain decomposition (such as the tiling scheme described here), each tile is required to exchange information with the neighboring tiles at each time step. This includes both grid information (edge values of fields and densities) and particle information (particles moving to/coming from another tile). Since each tile is self-contained, boundary information must be exchanged between all neighboring tiles---both tiles found within a single PE and those located across a PE--PE boundary. The exchange of boundary information introduces overhead not only from the MPI communications, but also from reassigning particles to different tiles within the same PE, with the latter overhead increasing as tile size decreases. If not optimized, these overheads can outweigh the benefit of the load balance algorithm, severely limiting its applicability.

In our implementation, boundary information between different PEs is packed into buffers and shared via MPI; tile boundary information within a single PE is instead referenced directly in shared memory. Whenever tiles are (re)assigned to PEs, each PE will gather and store information regarding which of its tile boundaries are internal and which are external. Each time step when boundary information is exchanged, boundaries are processed sequentially over the number of dimensions to avoid corner communications. When exchanging grid information, we loop over all tiles using parallel threads; local tile boundary values are copied directly in shared memory (no buffer required), and boundary values to be sent to other nodes are packed into one buffer per target node and sent via MPI. We then loop over all tiles with external boundaries (again using parallel threads) to copy in boundary values from the received communication buffers. Particles moving between tiles are processed in a similar manner. Each tile maintains buffers to store exiting particles; buffered particles moving to a local tile are referenced and unpacked, while particles moving to a different PE are packed and sent over MPI. Grouping MPI messages going from a local PE to the same target PE greatly improves performance (as opposed to using one message per tile boundary, for example) by reducing the number of messages and limiting the impact of communication latency, since message sizes are larger.

Using the tiling scheme with either of the space-filling curves will lead to a small increase in communication time when compared to a uniform partition. This is mainly due to an increased number of neighboring PEs (the corners of each PE are no longer aligned) that communicate smaller messages (i.e., communication is mostly latency-dominated). This increase in communication time can be offset by obviating the particle sorting step required in most PIC codes while still ensuring data locality and cache coherency (due to the small size of a tile).  Furthermore, the benefit of proper parallel load balance can greatly outweigh any penalties incurred from communication overhead for otherwise unbalanced runs.

We should also mention that the choice of tile size has a significant impact on the performance of the algorithm: smaller tile sizes improve data locality and load balance at the expense of a higher overhead from tile boundary management, as the number of particles (compared to the total particles in a tile) crossing boundaries will increase, as will the number of boundaries between PEs. On the other hand, larger tile sizes reduce the boundary management overhead but can interfere with data locality and hinder parallel load balancing by limiting the available resolution of the parallel partition. The optimal tile size will be problem-specific, as discussed in further detail in Sec.~\ref{sec:uniform-plasma}.

\end{subsection}

\section{Results} \label{sec:results}

To evaluate the performance of our dynamic load balance algorithm using tiles we will benchmark it against the baseline performance of \textsc{Osiris} using a static, regular spatial domain decomposition (referred to as ``no tiles'' in all figures). As mentioned earlier, these results depend on the type of problem and level of imbalance, as well as user choices such as tile size and dynamic load balance frequency. We will analyze simulations of a uniform warm plasma, an ambipolar diffusion problem, and a plasma wakefield accelerator. All simulations were performed on Haswell compute nodes of the Cori system at NERSC, each with two sockets, and each socket containing a 2.3~GHz 16-core Haswell CPU (Intel Xeon Processor E5-2698 v3) supporting 2 hyper-threads, leading to a total of 64 hardware threads per node.

\subsection{Uniform warm plasma} \label{sec:uniform-plasma}

\begin{figure}[t]
	\centering
	\captionsetup{width=1.0\linewidth}
	\includegraphics[width=0.95\linewidth]{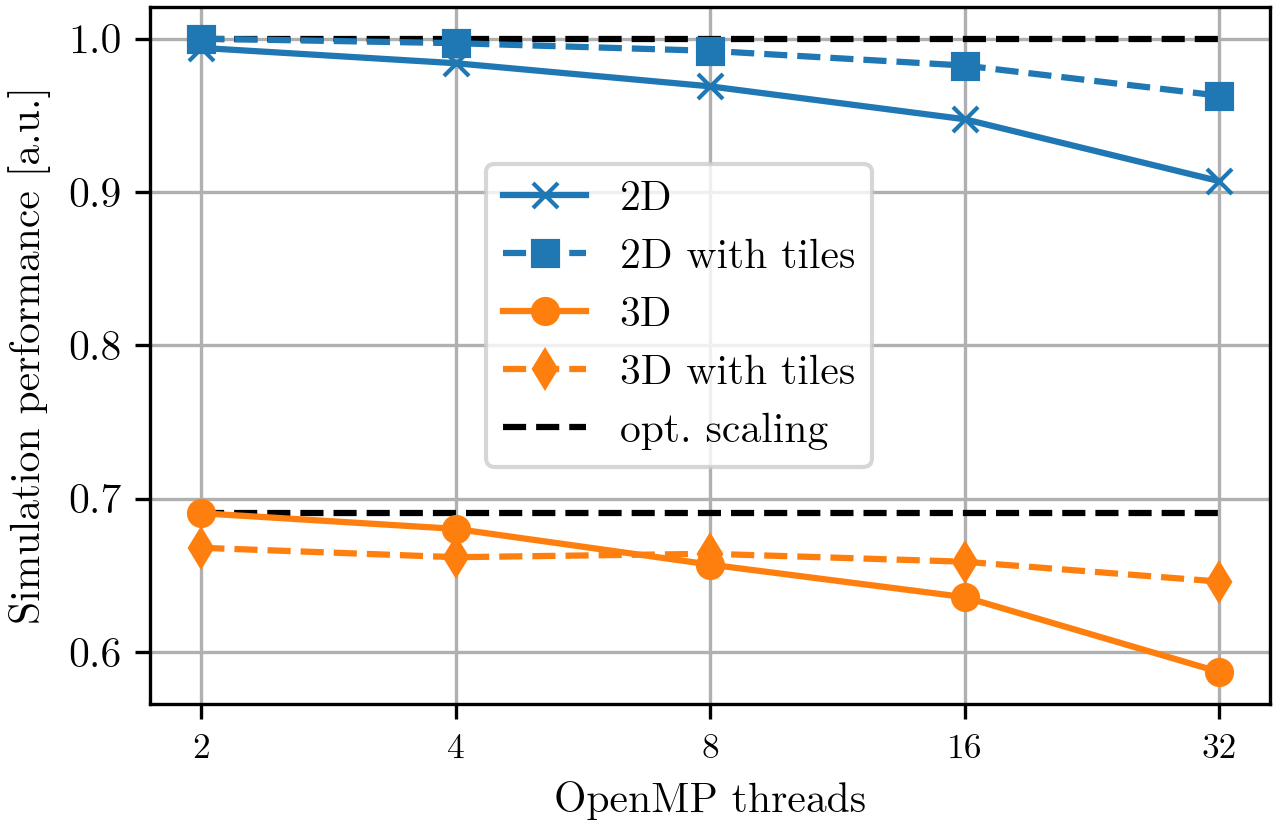}
	\caption{Thermal plasma simulation performance in 2D and 3D for varying number of threads, with (dashed line) and without (solid line) tiling algorithm. Tiles were 16 cells square in 2D and 8 cells cube in 3D. Number of MPI processes times number of threads per process was kept constant at 256. Performance is calculated as the number of particles pushed (including field solve and other elements) per second, normalized to the fastest run.}
	\label{fig:thermal-timing}
\end{figure}

A uniform warm plasma is a perfectly balanced problem that can be simulated using standard static spatial domain decomposition techniques with excellent parallel efficiency, since the load for every PE is uniform throughout the simulation. Simulating a warm plasma is thus ideal for benchmarking the overhead of the tiling algorithm compared to the default parallelization strategies---both pure MPI and MPI/OpenMP hybrid parallelization options are available in \textsc{Osiris} (see \cite{Fonseca2013} for details)---and optimizing tile size for best performance. We simulate a uniform warm plasma in 2D and 3D using four compute nodes, for a total of 128 cores (256 threads). The product of the number of MPI processes ($M$) and the number of threads per process ($N$) is kept constant at $M \times N=256$, for $N=2$, 4, 8, 16 and 32. The 2-D runs use a $1024^2$ cell grid with cell size $(0.0174 \, \mathrm{c}/\omega_p)^2$ and a total of $\sim 150$ million particles, and the 3-D runs use a $128^3$ cell grid with cell size $(0.0211 \mathrm{c}/\omega_p)^3$ and a total of $\sim 134$ million particles; both runs use a time step of $0.012\,\omega_p^{-1}$. Particle velocities were initialized from a thermal distribution with a proper thermal velocity of $u_{\mathrm{th}}=0.1 \, \mathrm{c}$.  All simulations were done using second-order (quadratic) interpolation for particles with periodic boundary conditions in all directions, and were run for 1000 time steps. Tile size was $16^2$ cells in 2D and $8^3$ cells in 3D. Figure~\ref{fig:thermal-timing} shows the performance of the code for the various cases.

The performance of the default hybrid parallelization in \textsc{Osiris} is shown to decrease with increasing thread number. This is to be expected as the shared-memory algorithm is required to do additional work (zeroing additional current arrays and reducing the results from all threads) with increasing thread count. Since all tiled simulations in this configuration have only ``light'' tiles (16 tiles for every thread), the new tiling algorithm does not suffer from this limitation and shows good scalability all the way up to 32 threads. The drop in performance is mostly due to the fact that not all routines have been shared-memory parallelized, as they do not represent a significant overhead for small thread counts.

The new algorithm is only outperformed by default \textsc{Osiris} for 3D geometry with small (2,4) number of threads. This is related to the overhead of moving particles between domains; default \textsc{Osiris} without tiles uses larger domains, so there will be fewer particles (compared to the particles in the domain) crossing boundaries. However, for larger thread counts the limitations of default \textsc{Osiris} outweigh this overhead, and the new algorithm performs much better.

To determine the optimal tile size, we repeated the above simulations for varied tile sizes and number of particles per cell. The latter parameter is important because it directly impacts the computational load of each tile. The number of particles per cell were varied between 1, 16, 64 and 144, and the tile sizes were varied by powers of 2, with values ranging from 8--64 cells on each side. Smaller values are not possible using second-order (quadratic) interpolation, as a minimum of 5 cells is required. Table~\ref{table:thermal-tile-size} summarizes our results. We found that the optimal tile side length varied between 16--64 cells in 2D and 8--16 cells in 3D. These results are consistent with our expectations: when the amount of computation per tile is larger (as is the case with higher numbers of particles per cell), the overhead of moving particles between tiles has a smaller impact, and the benefits of smaller tiles in terms of data locality lead to higher performance. For smaller numbers of particles per cell this is no longer the case, and larger tiles perform better.

\begin{table}[t]
	\centering
	\captionsetup{width=1.0\linewidth}
	\begin{tabular}{ P{0.17\linewidth}|P{0.115\linewidth} P{0.07\linewidth} P{0.08\linewidth}|P{0.07\linewidth} P{0.07\linewidth}}
		 & \multicolumn{3}{c|}{2D} & \multicolumn{2}{c}{3D} \\
		\hline
		Particles per cell & 1 & 16 & 64+ & 1 & 8+ \\
		\hline
		Optimal tile size & 32--64 & 32 & 16 & 16 & 8 \\
	\end{tabular}
	\caption{Optimal tile sizes (cells on each side) in 2D and 3D for a thermal plasma with varied particles per cell. Decreasing the tile size introduces more overhead but improves cache locality, which is important for simulations with many particles per cell.}
	\label{table:thermal-tile-size}
\end{table}


%


\subsection{Ambipolar diffusion}

Ambipolar diffusion is of particular importance to a large range of physics scenarios, such as laser-solid interactions and inertial confinement fusion \cite{Chen1995,Lindl1995DevelopmentGain,Betti2016Inertial-ConfinementLasers}. This is a particularly difficult problem to simulate using spatial domain decomposition as the plasma, which is initially confined to a small region of space, expands into vacuum (or near vacuum). We will use this problem to test the effectiveness of our algorithm under extreme load imbalance. We perform a simulation of the expansion of an electron-ion plasma undergoing ambipolar diffusion in 3D, using the various parallelization strategies discussed in Sec.~\ref{sec:methodology}. We start with a constant-density sphere of electrons and ions, where the electrons have a temperature of 130~keV and the ions are cold.  The sphere has radius 1.4~$c/\omega_p$ and is stationed in the center of a cubic periodic box of side length 16~$c/\omega_p$ with $160^3$ cells. The electrons quickly diffuse outward, but are soon slowed by the ambipolar space charge fields of the ions.  The ions are in turn pulled out and start to expand, which slows down the expansion of the electrons. Afterwards the electrons will again start to diffuse, leading to an oscillatory expansion of both species.

\begin{figure*}
	\centering
	\captionsetup{width=1.0\linewidth}
	\includegraphics[width=0.85\linewidth]{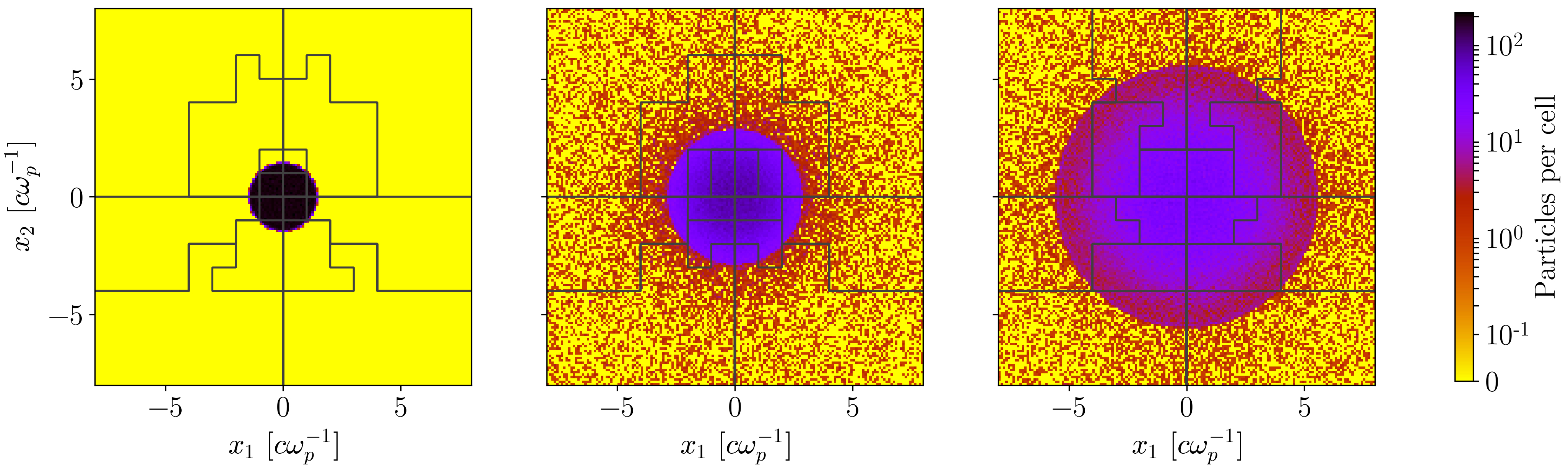}
	\caption{The 16 PE subdivisions overlaying particle density of the ambipolar diffusion problem in 2D at various times.  Cell weight is 2.0 for this case.}
	\label{fig:ambipolar-topologies}
\end{figure*}

Figure~\ref{fig:ambipolar-topologies} shows the combined electron-ion particle density for an analogous 2-D simulation at times 0.0, 152.71, and 299.77~$\omega_p^{-1}$ after the beginning of the expansion, where $\omega_p$ is the electron plasma frequency before expansion. The domains of the 16 PEs overlay the density; density slices along any dimension in 3D show similar behavior, but the parallel domain decomposition can be better visualized and understood in 2D. For the first half of the 3-D simulation, about 70\% of the particles are contained within a sphere of radius 2.5~$c/\omega_p$, or just 1.6\% of the entire simulation volume.  By the end of the simulation, 70\% of the particles are contained within a sphere of radius 5~$c/\omega_p$ (13\% of the total volume). Since the particles largely reside in the center of the simulation box, a traditional spatial domain decomposition with uniform partition sizes will only allow good parallel load balance for up to 2~PEs per dimension (8~PEs total), and will show severe imbalance if the number of PEs is increased to 4 or more. This simulation is also challenging for our tile-based approach: using a total of 4096 cube-shaped tiles with 10 cells to a side, 70\% of the particles are contained within just 56 tiles for the first third of the simulation, steadily increasing to 432 tiles by the end of the simulation.

\begin{figure*}
	\centering
	\captionsetup{width=1.0\linewidth}
	\includegraphics[width=0.8\linewidth]{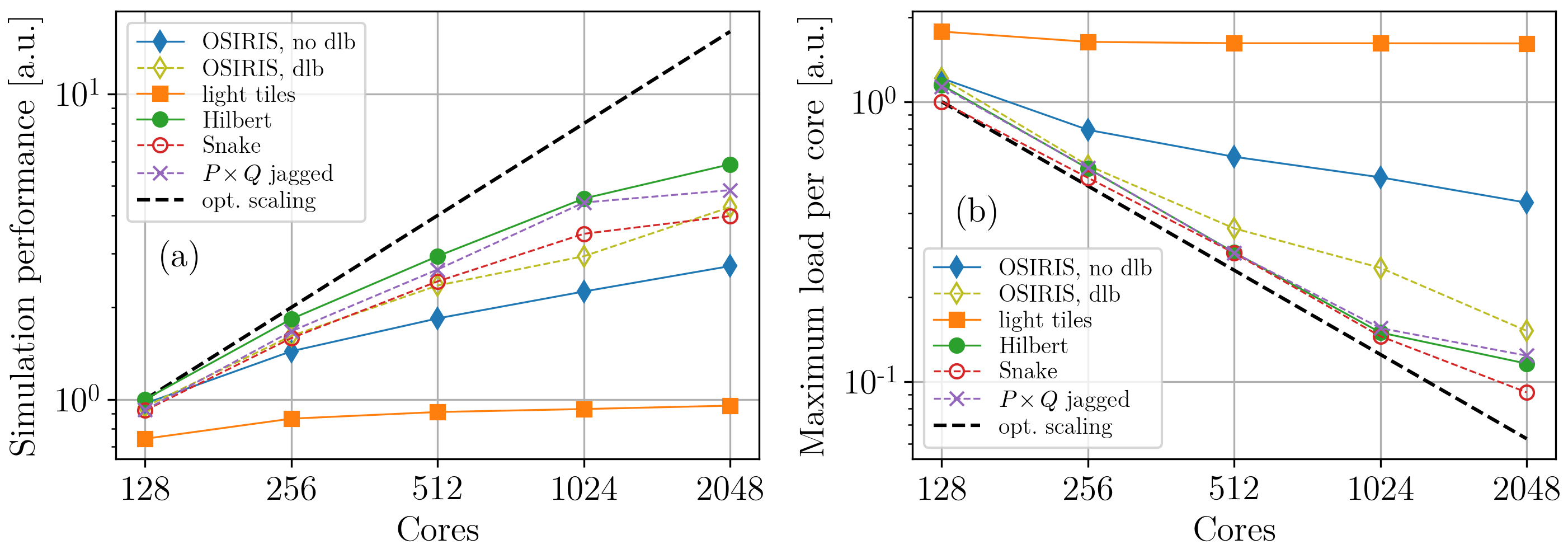}
	\caption{Strong scaling test of a 3-D ambipolar diffusion simulation. Performance in (a) is calculated as the number of particles pushed (including field solve and other elements) per second, and load in (b) is calculated as the maximum time any core spent in the advance-deposit routine, both normalized to the fastest run with 128 total cores. All cases without tiles are performed with 16 threads per MPI process, with tiled runs using an optimal combination of MPI processes/threads (between 2--16).}
	\label{fig:ambipolar-adv-dep}
\end{figure*}

We test the strong scaling parallel speedup of this simulation, keeping the problem size constant and running on 128 to 2048 cores for various cases: the default hybrid MPI/OpenMP algorithm without (``\textsc{Osiris}, no dlb'') and with (``\textsc{Osiris} dlb'') the previously implemented dynamic load balance~\cite{Fonseca2013}, light tiles only with dynamic load balance using the Hilbert space-filling curve (``light tiles''), and light/heavy tiles with dynamic load balancing using all three schemes (``Hilbert,'' ``Snake,'' and ``$P\times Q$ jagged''). We choose to use the maximum number of threads per MPI process without hyperthreading (16) as this is the most favorable scenario for the default hybrid algorithm. Figure~\ref{fig:ambipolar-adv-dep} shows the results in terms of (a)~overall simulation performance and (b)~load per core, normalized to the fastest simulation with 128 cores.

The default hybrid algorithm without dynamic load balancing is unable to scale due to load imbalance issues. Figure~\ref{fig:ambipolar-adv-dep} also shows that the scalability of the light tiles algorithm is severely limited as a result of its inability to parallelize within each tile, and in fact the light tiles algorithm is the slowest algorithm at all core counts. In a worst-case scenario, during the first third of the simulation the 56 computationally expensive tiles could be spread across just 4 PEs (64 cores) since the algorithm groups together tiles close in space. With only light tiles each tile can only be processed by a single core, leading to significant performance degradation that worsens with increasing core counts.

This limitation in parallel scalability when using only light tiles is overcome by our new algorithm that uses a combination of light and heavy tiles. We implement an additional level of parallelization within each tile, which allows a PE to efficiently devote all of its cores to process a single computationally expensive tile. As seen in Fig.~\ref{fig:ambipolar-adv-dep}, our algorithm (a) maintains much better overall parallel scalability compared to any other method and (b) achieves near ideal load balance for up 1024 cores. At all core counts, the combination of heavy/light tiles always outperforms the use of light tiles only, being $\sim 2$ times faster for 128 cores and $\sim 6$ times faster for the largest core count. The average number of heavy tiles over the course of the simulation steadily increases with thread count, ranging between 0.5\% and 8.2\% of all tiles from 128 to 2048 cores, respectively. The small drop in (b) from ideal scalability at 2048 cores has to do with the problem/tile size: at this core count we will have on average only 2 tiles per PE, which does not allow for effective load balancing. Reducing the tile size may improve performance for this case.

When comparing the various load-balancing schemes, using the Hilbert space-filling curve was consistently the fastest, and with 2048~cores it was 1.2, 1.5, and 1.4 times faster than the $P\times Q$ jagged scheme, snake space-filling curve scheme, and previous dynamic load balance, respectively.  An overall speedup of a factor of 2.2 was gained compared to original \textsc{Osiris} without dynamic load balance.  Though the snake space-filling curve consistently gave excellent load balance, extra overhead from MPI communications due to the shapes of PE boundaries ultimately caused those runs to be slower than the other topologies.



\subsection{Plasma Wakefield Acceleration} \label{sec:PWFA}

Since its inception in the seminal paper by Tajima and Dawson \cite{Tajima1979LaserAccelerator}, the field of plasma wakefield acceleration has been an active area of intense research. Plasma based acceleration (PBA) \cite{Joshi2018PlasmaII} is an accelerator scheme which uses an intense electron bunch---particle wakefield acceleration (PWFA)~\cite{Chen1985AccelerationPlasma}---or a laser---laser wakefield acceleration (LWFA)---to accelerate particles. As the beam propagates through the plasma, the electrons are pushed away from the beam to leave an exposed ion column, resulting in a region of space supporting large electric fields that travels at nearly the speed of light. Particles can be injected and trapped from the background plasma through various mechanisms~\cite{Lu2007GeneratingRegime,Suk2001PlasmaTransition,Oz2007Ionization-inducedWakes,Xu2017HighRegime}, then accelerated to high energies. The excitation of the wake by the drive beam is strongly nonlinear, as can be the evolution of the driver and the injected and accelerated beams, making PIC simulations the tool of choice for modeling these scenarios.

Simulations of PWFA naturally contain regions of very high particle density---the injected particle bunch and beam driver---that are dynamic in nature, surrounded by regions of relatively low-density background plasma. These density distributions can result in severe load imbalance, for which finding a single PE decomposition that balances load for the entire simulation may be near impossible. Effective dynamic load balancing of the simulation is therefore crucial for efficient numerical modeling of PWFA, particularly when using large core counts. We test the performance of our dynamic load balancing algorithm with a PWFA scenario similar to that studied by Dalichaouch et al.~\cite{Dalichaouch2020GeneratingDriver}. The driving beam is initialized with a radius of $\sigma_r = 2.1 \, \mathrm{c}/\omega_p$ and then evolves self-consistently, focusing down to a radius of $\sigma_r = 0.2 \, \mathrm{c}/\omega_p$. When initializing a beam, to provide good statistics we use a fixed number of particles per cell and variable weights on the particles to vary the density. The beam evolution leads to a large load imbalance due to the large number of simulation particles concentrated in a small cell volume. Additionally, the accelerating structure formed with particles from the background plasma will show a density spike at the back of this structure that can be several orders of magnitude larger than the background density, creating a second load imbalance region.

\begin{figure}[t]
	\centering
	\captionsetup{width=1.0\linewidth}
	\includegraphics[width=1.0\linewidth]{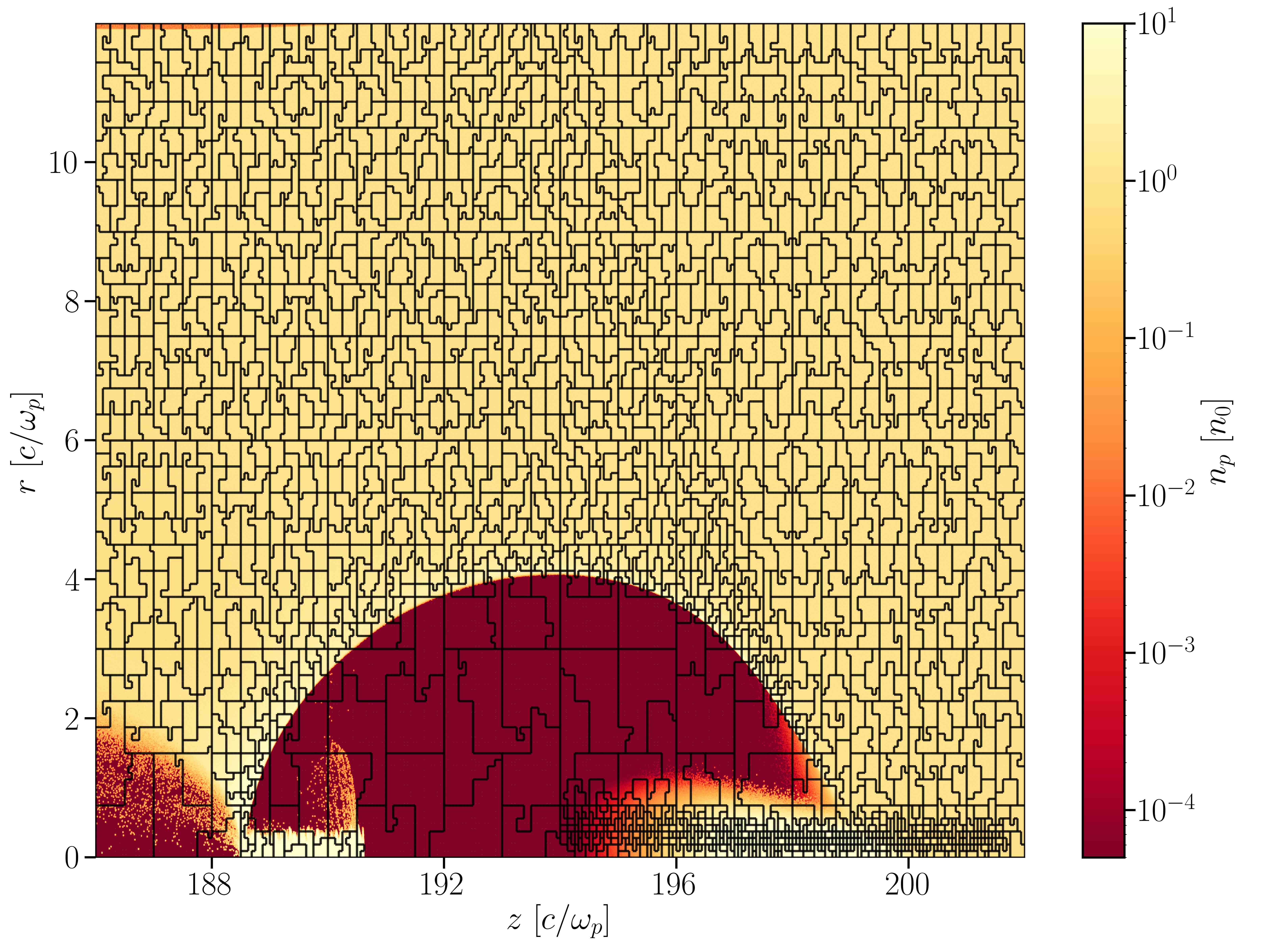}
	\caption{MPI subdivisions (black lines) overlaying electron density for a quasi-3D PWFA simulation with 2048 MPI processes.}
	\label{fig:rz-wakefield}
\end{figure}

To illustrate the complexity of finding an optimal domain decomposition, we show in Fig.~\ref{fig:rz-wakefield} a snapshot from an analogous simulation using the quasi-3D version of \textsc{OSIRIS} that uses a geometry (2-D $r$-$z$ in space, azimuthal expansion in $\theta$) \cite{Davidson2015ImplementationOSIRIS}. Load balance was performed using the Hilbert space-filling curve and a cell weight $C=1$ for 2048 PEs; the simulation size was $2048\times1536$ cells, with $256\times128$ tiles each of size $8\times12$ cells. This snapshot is about halfway through the simulation when the particle driver has focused tightly near to the axis, and the figure shows an overlay of node boundaries with electron density.  Note the large concentration of small domains near the beam driver, bubble sheath and injected particles, and the large domains in the near-vacuum region. Similar behavior occurs for 3D simulations, but the visualization of these structures is cluttered and brings little insight.

We tested the efficiency of our algorithm on 3D simulations of this PWFA scenario by performing a set of simulations with different parallelization options. All simulations were performed on 64 compute nodes with the architecture described above (2048 cores total) and varying numbers of MPI processes/threads, for the default hybrid parallelization without load balance and for the dynamic light/heavy tiles parallelization. The simulation size was $512\times768\times768$ cells, and the tile size was $16 \times 12 \times 12$ cells for the tile-based simulations. The simulation was run for a total of 12800 time steps, with dynamic load balancing performed at every 40 time steps. For load calculations we chose a cell weight parameter of $C=0.5$.

\begin{figure}[t]
	\centering
	\captionsetup{width=1.0\linewidth}
	\includegraphics[width=0.95\linewidth]{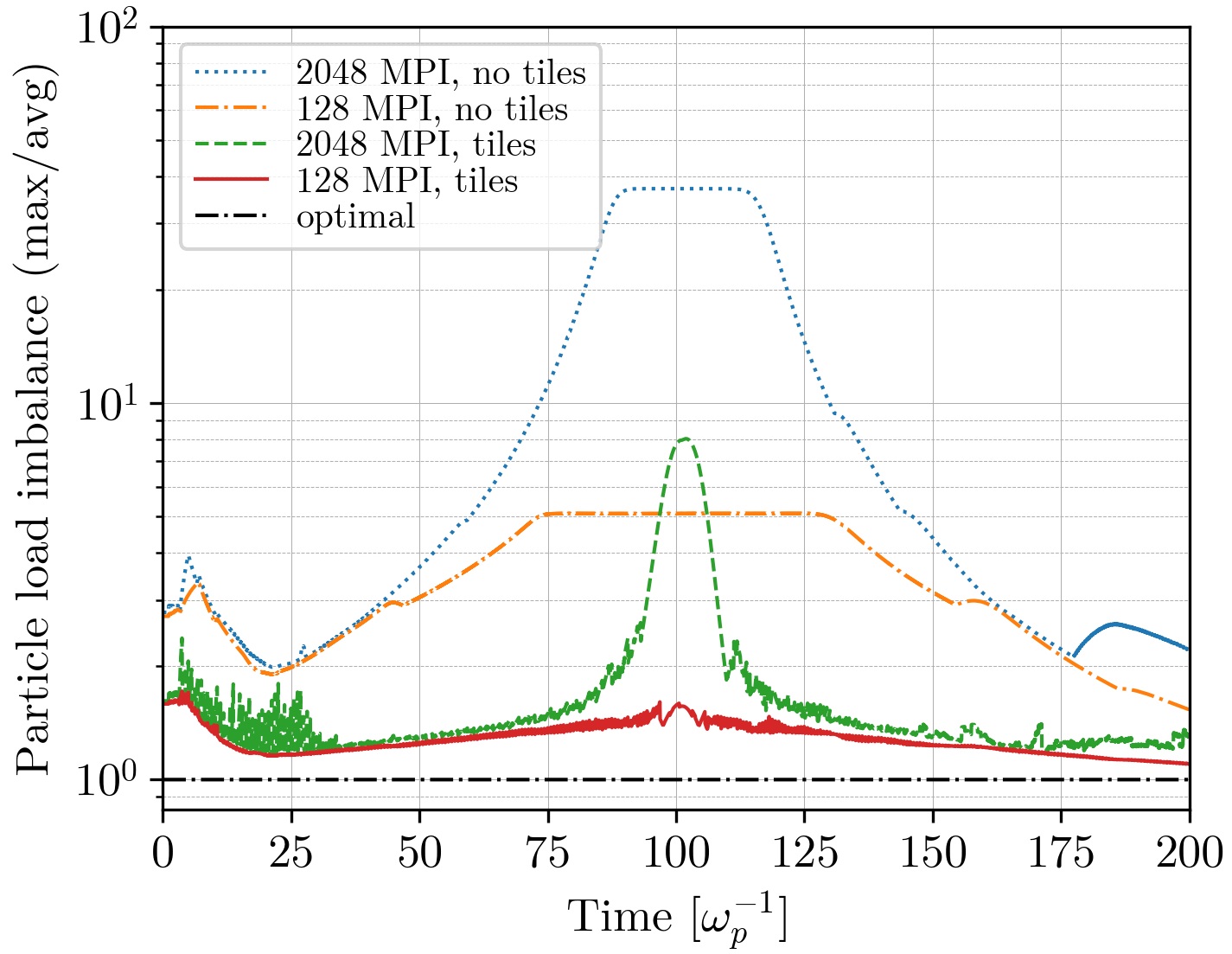}
	\caption{Particle load imbalance (maximum/average load) of a 3-D PWFA simulation using 2048 total cores with and without dynamic load balancing using tiles for two different MPI/OpenMP configurations. The dot-dashed line shows an ideal load imbalance of 1.0.}
	\label{fig:3d-wakefield-node-load}
\end{figure}

Figure~\ref{fig:3d-wakefield-node-load} shows the evolution of the parallel particle load imbalance over the course of a simulation using 2048 total cores with two MPI/OpenMP configurations (see Table~\ref{table:3d-wakefield-timings} for more detail) as the ratio between the maximum computational load on a single PE and the average computational load across all PEs. The load imbalance is shown for both the static uniform partition (no tiles) and the dynamic partition (tiles). As shown in the plot, the high-density regions near the propagation axis lead to a severe load imbalance about halfway through the simulation. For 2048 processes each with 1 thread, the imbalance reaches a peak (average) value of 37.1 (10.6) without dynamic load balancing. This result can be improved through the use of the hybrid MPI/OpenMP algorithm~\cite{Fonseca2013} that smears out the load imbalance by allowing for large domains to be assigned to each PE, leading to a peak (average) imbalance value of 5.1 (3.5) at 128 processes each with 16 threads. It should, however, be noted that this large number of threads leads to some performance degradation due to overhead, as described in Secs.~\ref{sec:shared-memory} and \ref{sec:uniform-plasma}. To further reduce the parallel load imbalance we must use the dynamic tile load balance. Using this algorithm we can lower the peak (average) value of the imbalance down to 8.0 (1.7) with 2048 processes, as shown in Fig.~\ref{fig:3d-wakefield-node-load}. Furthermore, using our heavy/light tile parallelization we can assign multiple threads to the heavy tiles, achieving a peak (average) load imbalance of just 1.7 (1.3) with 128 processes.  Note that for this run, nearly all tiles were processed as light except for an average of 175 heavy tiles near the focus of the beam driver during the middle 10\% of the simulation time.

Finally, we compare the total simulation times for the different configurations of MPI processes and OpenMP threads for the static uniform partition and the dynamic tiles partition. Table~\ref{table:3d-wakefield-timings} shows the results, with time values normalized to the fastest simulation (512 MPI processes with 4 OpenMP threads and dynamic load balance). As shown in the table, performance of the static partition (no tiles) can be improved by over a factor of 2 simply by increasing the number of threads. However, the tile-based algorithm maintains good performance for all thread counts and always outperforms the static partition, being 2.18 times faster between the fastest runs. Cumulative load balance is roughly the same for tile-based runs with 4, 8, and 16 OpenMP threads, but the overhead discussed in Sec.~\ref{sec:ub} is slightly larger with 16 threads. The overhead associated with the dynamic load balance effort, however, was found to be only $\sim 1 \%$ of the total simulation time in all cases. We also see that the timings for the tile-based algorithm vary less than $10 \%$ with varying number of threads, which is to be expected, as the computation should scale well with number of threads. This small variation highlights the versatility of the algorithm in terms of efficiency on various architectures supporting different numbers of threads.

\begin{table}[t]
	\centering
	\captionsetup{width=1.0\linewidth}
	\begin{tabular}{ P{0.18\linewidth} P{0.173\linewidth} P{0.211\linewidth} P{0.211\linewidth} }
		MPI processes & OpenMP threads & Time [a.u.] no tiles & Time [a.u.] tiles \\
		\hline
		2048 & 1 & 4.63 & 1.19  \\
		1024 & 2 & 4.55 & 1.03  \\
		512 & 4 & 2.45 & 1.00 \\
		256 & 8 & 2.57 & 1.03 \\
		128 & 16 & 2.18 & 1.10  \\
	\end{tabular}
	\caption{Total simulation time with and without tiles (and dynamic load balancing) for the 3-D PWFA simulation.  The total number of cores was kept constant (2048), only varying the number of OpenMP threads and MPI processes.}
	\label{table:3d-wakefield-timings}
\end{table}

\section{Conclusion}

Many of today's frontier problems in plasma physics necessitate fully kinetic simulations, which can in many cases only be achieved through particle-in-cell simulations. Although the PIC algorithm has been quite successful in addressing many kinetic plasma problems (and can be extended straightforwardly with additional physics models), it is numerically expensive, often requiring large-scale, massively-parallel computational resources. However, the traditional parallelization of the PIC algorithm is susceptible to parallel load imbalance. In problems such as plasma-based acceleration, laser-solid interactions, and magnetic reconnection with pair production, simulation particles may accumulate in small regions of space, leading to an uneven computational load. Dynamic load balancing across distributed-memory processing elements (PEs) and the efficient use of shared-memory cores are therefore essential to perform large-scale simulations of these physics problems.

In this paper we presented a novel hybrid parallelization strategy for the PIC algorithm that combines two different shared-memory parallelization algorithms, achieving excellent performance even for simulations with extreme imbalance. Our algorithm uses small, regularly spaced, self-contained regions of the simulation space that we refer to as tiles. These tiles contain all particle and grid quantities for a particular region of space and are dynamically traded between all PEs at chosen intervals to maintain computational load balance. Unique to our tile-based implementation of dynamic load balancing is the ability to assign either one or multiple threads to each tile depending on computational weight. This is especially useful for simulations with localized high-density regions for which the load-balanced domain decomposition may result in a PE having a single tile containing a majority of particles, or for simulations where a given PE has fewer tiles than threads. Furthermore, the ability to assign multiple threads to each tile allows us to use larger tiles, which can reduce the overhead of the tiling algorithm while still maintaining good load balance and high performance.

Our algorithm was shown to perform well for balanced simulations: it was on par with or faster than the traditional parallelization algorithm on the same hardware and scaled better with increasing thread count, with an overhead of less than $\sim$5\% for the highest thread count. Our algorithm gave a speedup of more than a factor of 2 compared to simulations without dynamic load balance of an expanding plasma undergoing ambipolar diffusion, a critically difficult problem to parallelize with 70\% of the particles contained within only 13\% of the simulation volume. In particular, it was $\sim 6$ times faster than running the simulation using only 1 thread per tile and attained near-ideal load balance for 8 times as many cores compared to running without tiles. We also analyzed the performance of our algorithm on plasma accelerator simulations and verified a speedup of over a factor of 2 when compared to performance without dynamic load balance, with the dynamic load balance itself only taking about 1\% of the total simulation time. Even greater speedups are expected for larger simulations, where shared-memory thread number is small compared to total computing resources. Our results also show that the performance of the algorithm does not vary significantly with different combinations of PE and thread numbers.

This algorithm was tested on a CPU architecture using MPI / OpenMP for distributed / shared memory parallelism, and can be straightforwardly extended to other architectures and programming models, such as GPUs using CUDA, or other architectures using Intel oneAPI. It can also be combined with vectorized versions of the PIC algorithm, efficiently exploiting all parallelism levels available in present and near-future HPC systems and opening new avenues for the numerical simulation of kinetic plasmas.

\section*{Acknowledgements}
This work was supported in parts by the US National Science Foundation [grant numbers ACI-1339893, 1806046]; the US Department of Energy [grant numbers DE-NA0003842, DE-SC0019010]; Lawrence Livermore National Laboratory [subcontract B634451]; Fundação para a Ciência e Tecnologia, Portugal [grant number PTDC-FIS-PLA-2940-2014]; and European Research Council [ERC-2015-AdG, grant number 695008]. The simulations were performed on Haswell CPUs (Intel Xeon Processor E5-2698 v3) of the Cori system at NERSC.

\bibliographystyle{elsarticle-num-names}
\bibliography{references}

\end{document}